\documentclass[12pt,a4paper]{article}            
 \usepackage[skins,theorems]{tcolorbox}
\tcbset{highlight math style={enhanced,
  colframe=red,colback=white,arc=0pt,boxrule=1pt}}
  \usepackage[bookmarksopen, bookmarksnumbered, bookmarksopenlevel=2]{hyperref}
  \usepackage{tikz}
  \usepackage{tikz-3dplot}
 \usetikzlibrary{calc}
 \usetikzlibrary{decorations} %
 \usepackage[UKenglish]{babel}
 \usepackage[toc,page]{appendix}
 \usepackage{amsmath}
 \usepackage{amssymb}
 \usepackage{graphicx}
 \usepackage{hhline}
 \usepackage{tikz-feynman}
 \usepackage[bf]{caption}
\usepackage{cite}
\usepackage[vcentermath]{youngtab}
\usepackage{geometry}
\usepackage{slashed}
\usepackage{color}
\usepackage{stackrel}
\usepackage{tikz-cd} 
\usepackage{mathtools}
\usepackage{cancel} 
\usepackage{multirow}
\usepackage[margin=0pt,font=small,labelfont=normalfont,skip=22pt]{subcaption}

\usepackage{empheq}
\usepackage{arydshln}

\newcommand{\bqa}{\begin{eqnarray}}
\newcommand{\eqa}{\end{eqnarray}}

\newenvironment{eqn*}{\begin{equation*}\begin{aligned}}{\end{aligned}\end{equation*}\noindent}
\hypersetup{
    pdftitle={},
    pdfauthor={},
    pdfsubject={}
}
\numberwithin{equation}{section}
\numberwithin{table}{section}

\makeatletter

\newcommand{\be}{\begin{equation}}
\newcommand{\ee}{\end{equation}}
\newcommand{\beq}{\begin{equation}}
\newcommand{\eeq}{\end{equation}}
\newcommand{\ba}{\begin{aligned}}
\newcommand{\ea}{\end{aligned}}

\newcommand{\bea}{\begin{eqnarray}}
\newcommand{\eea}{\end{eqnarray}}

\newcommand{\cN}{\mathcal{N}}

\newcommand\bi{\begin{itemize}}
\newcommand\ei{\end{itemize}}

\def\Tr{\mathop{\mathrm{Tr}}\nolimits}

\def\unit{{1\kern-.65ex {\rm l}}}
\def\1{{1\kern-.65ex {\rm l}}}

\newcount\hour \newcount\minute
\hour=\time \divide \hour by 60
\minute=\time
\def\now{%
\ifnum \hour<13
  \ifnum \hour=0 \advance \hour by 12 \number\hour:\else \number\hour:\fi%
     \ifnum \minute<10 0\fi%
     \number\minute%
\ A.M.%
\else \advance \hour by -12 \number\hour:%
  \ifnum \minute<10 0\fi%
  \number\minute%
  \ P.M.%
\fi%
}

\makeatother

\begin{document}

\begin{titlepage}
\begin{center}
\rightline{\small }

\vskip 15 mm

{\large \bf
Bounds on Species Scale and the Distance Conjecture
} 
\vskip 11 mm

Damian van de Heisteeg,$^{1}$ Cumrun Vafa,$^{2}$ Max Wiesner$^{1,2}$

\vskip 11 mm
\small ${}^{1}$ 
{\it Center of Mathematical Sciences and Applications, Harvard University,\\ Cambridge, MA 02138, USA}  \\[3 mm]
\small ${}^{2}$ 
{\it Jefferson Physical Laboratory, Harvard University, Cambridge, MA 02138, USA}

\end{center}
\vskip 17mm

\begin{abstract}
The species scale $\Lambda_s\leq M_{\rm pl}$ serves as a UV cutoff in the gravitational sector of an EFT and can depend on the moduli of the theory as the spectrum of the theory varies.
We argue that the dependence of the species scale $\Lambda_s (\phi)$ on massless (or light) modes $\phi^i$ satisfies $M_{\rm pl}^{d-2} |\Lambda_s'/\Lambda_s|^2< \mathcal{O}(1)$.  This bound is true at all points in moduli space including also its interior.  The argument is based on the idea that the short distance contribution of massless modes to gravitational terms in the EFT cannot dramatically affect the black hole entropy.  Based on string theory arguments we expect the $\mathcal{O}(1)$ constant in this bound to be equal to ${1\over {d-2}}$ as we approach the boundary of the moduli space. However, we find that the slope of the species scale can approach its asymptotic value from above as we go from interior points to the boundaries, thereby implying that the constant in the bound must be larger than ${1\over {d-2}}$. The bound on the variation of the species scale also implies that the mass of towers of light modes cannot go to zero faster than exponential in field distance in accordance with the Distance Conjecture. 
\end{abstract}

\vfill
\end{titlepage}

\newpage

\tableofcontents

\setcounter{page}{1}

\section{Introduction}
The species scale, $\Lambda_s$, was introduced \cite{Dvali:2007hz,Dvali:2010vm, Dvali:2009ks,Dvali:2012uq} to signify a new cutoff in the gravitational sector which in the presence of a large number of light particles is smaller than the Planck scale. On the other hand the Distance Conjecture \cite{Ooguri:2006in} implies that such towers of light particles emerge at the boundaries of moduli space.  Thus it is expected that $\Lambda_s$ depends on the massless moduli of the theory.  This dependence has been studied in \cite{Long:2021jlv,vandeHeisteeg:2022btw} including at interior points of the moduli space.

The aim of this paper is to put a bound on the variation of $\Lambda_s$ as a function of the moduli fields.   In particular we will argue that $\Lambda_s$ as a function of a canonically-normalized light scalar field, $\phi$, satisfies
\begin{equation} \label{inequality}
\left|{\Lambda'_s(\phi)\over \Lambda_s(\phi)}\right |^2 <{c\over M_{\rm pl}^{d-2}}\,,
\end{equation}
for some (possibly dimension-dependent) $\mathcal{O}(1)$ number $c$. We derive this bound by studying the structure of the gravitational effective action obtained upon integrating out the massive modes.  Using arguments about black hole entropy, we then argue that this structure should not be affected by further integrating out the short distance modes of the moduli fields, $\phi$, which leads to the above inequality.

The inequality \eqref{inequality} in particular implies that the species scale cannot vary any faster than exponentially in the distance on moduli space. Using the power law relation between the mass of the tower of light particles and the species scale we then argue that asymptotically also the mass of the tower cannot go to zero any faster than exponential.  This is consistent with the Distance Conjecture which demands the existence of such a tower whose mass vanishes exponentially as we approach the boundaries.

The organization of this paper is as follows:  In section~\ref{sec:review} we recall aspects of the Distance Conjecture and the behavior of the species scale as we approach the boundaries of the moduli space.  In section~\ref{sec:boundderivation} we argue for the inequality \eqref{inequality} by studying the effective gravitational action.  In section~\ref{sec:typeII} we specialize to Type II compactifications on Calabi--Yau threefolds and check the validity of our bound in these setups.  In section~\ref{sec:conclusions} we present some concluding remarks.

\section{Species Scale and the Distance Conjecture}\label{sec:review}
In this section we review aspects of the species scale \cite{Dvali:2007hz,Dvali:2010vm, Dvali:2009ks,Dvali:2012uq} and its relation to the Distance Conjecture \cite{Ooguri:2006in}.
For a review of these topics within the broader context of the Swampland program see \cite{Palti:2019pca,vanBeest:2021lhn, Agmon:2022thq}. 

Whenever dealing with black holes in the context of gravity, one typically does not consider black holes for which the curvature at the horizon is as big as the Planck scale.  The reason for this is that one expects corrections to the Einstein action which involve higher-dimension operators (like powers of the Riemann tensor and various derivatives) and that, compared to the Einstein term, the terms involving these operators are suppressed by additional powers of $M_{\rm pl}$.  For example, each additional factor of $R$ in the action will be suppressed by an additional factor of $1/M_{\rm pl}^{2}$.  This in particular means that the higher-order terms are comparable to the Einstein term when $R
\sim M_{\rm pl}^2$. Therefore, when discussing configurations with Planckian curvatures, we cannot ignore all the higher-order terms and thus the EFT involving just the Einstein term breaks down.  This in particular necessitates considering black holes whose radius is at least $R_{\rm min}>M_{\rm pl}^{-1}$.

However, it has been argued \cite{Dvali:2007hz} that whenever there exists a large number of species of light fields, the suppression factors of the higher-dimension operators should not be just given by the Planck mass.  The reason for this is that in the presence of a large number of light species, the smallest black hole which we can hope to describe within the EFT should be bigger.  Otherwise the entropy of the smallest black hole would be of order one, whereas it should be at least of the order of the number of light species.  In other words, we expect the radius of the smallest black hole describable using the EFT to be at least $R_{\rm min}> N^{1\over {d-2}} M_{\rm pl}^{-1}$,  where $N$ is the ``number of light species''.
In particular this implies that the effective suppression factor in the EFT for the higher-derivative terms should instead of $M_{\rm pl}$ involve a smaller scale given by
\begin{equation}\label{defspeciesscale}
    \Lambda_s=M_{\rm pl} \,N^{-1\over d-2}\,,
\end{equation}
which is called the species scale \cite{Dvali:2007hz}.  To be more precise about what one means by the number of light species, we need to explain how to determine $N$ and $\Lambda_s$ which are mutually defined as follows.  One may naively expect that $N$ counts the number of states lighter than $\Lambda_s$.  However this is not precisely correct and yields an incorrect answer in case the light species arises as excitations of a light string. Instead, $N$ is the logarithm of the number of ways to obtain the minimum-sized black hole of radius $R\sim 1/{\Lambda_s}$ and total mass $m_{BH}\sim M_{\rm pl}(M_{\rm pl}/\Lambda_s)^{d-3}$ from combination of states lighter than $m_{BH}$ which should of course agree with the black hole entropy $S\sim R \cdot m_{BH}$.   
As explained in \cite{Agmon:2022thq}, with this definition the species scale $\Lambda_s$ becomes $M_s$ in the light string case and the higher-dimensional Planck mass in the KK case.  

Viewing the species scale as a UV cutoff we expect that each extra factor of the curvature $R$ in the EFT should be suppressed by an additional power of $1/\Lambda_s^2$, instead of $1/M_{\rm pl}^2$.  This means that the general form of the action for the gravitational sector of the theory coupled to massless scalars $\phi$ is expected to be
\begin{equation}\label{generalform}
    S = \int \mathrm{d}^dx \, \sqrt{-g} \left[\frac{M_{\rm pl}^{d-2}}{2} \left(R +\sum_n  \,\frac{\mathcal{O}_n(R)}{\Lambda_s^{n-2}(\phi)}+\dots  \right) - \frac{1}{2}(\partial \phi)^2+\dots\, \right]. 
    \end{equation}
where $\mathcal{O}_n(R)$ is a gravitational operator of dimension $n$ and the $\dots$ in the above equation include the terms involving all the light fields with mass less than $\Lambda_s$. Even though in a given theory not all possible operators $\mathcal{O}_n(R)$ may appear, in general we expect the coefficients of $\mathcal{O}_n(R)$ to be $\mathcal{O}(1)$ numbers.
Indeed it was argued in \cite{Heckman:2019bzm} that there can at most be a finite number of fine-tunings in the higher-derivative terms for a theory with a finite number of massless modes.
The above form of the action explains why the naive formula for black hole entropy will begin to fail for black holes smaller that $1/\Lambda_s$, because the curvature terms for these black hole are of order of $\Lambda_s^2$ and so all the higher order terms become relevant and can modify the naive expectation based on the Einstein-Hilbert action dramatically.

As one varies the scalar field vevs, $\langle \phi\rangle$, the number of light degrees of freedom changes and thus $\Lambda_s$ varies.
In particular, as we approach the boundaries of moduli space we expect, according to the Distance Conjecture \cite{Ooguri:2006in}, a tower of light states to appear whose mass goes to zero exponentially fast.  Let us briefly review this expected behavior, and use it to predict the behavior of $\Lambda_s$ as we approach the boundaries of moduli space.

\subsection{Asymptotic Behavior of the Masses}
String theory suggests that as we approach the boundaries of moduli space, we get a light tower of states, which are weakly coupled and become the basic ingredients of a dual description. The Distance Conjecture \cite{Ooguri:2006in}, which captures this duality phenomenon, posits that as we move a large distance in moduli space, there exists a tower of light states, whose mass scales as
\begin{equation}\label{distanceconj}
    m\sim {\rm exp}(-\alpha \phi )\,,
\end{equation}
where $\alpha \sim \mathcal{O}(1)$ in Planck units and the modulus $\phi$ is canonically normalized.  For simplicity of notation we sometimes set $M_{\rm pl}$ in the $d$-dimensional theory to 1.  The discussion below follows that of \cite{Agmon:2022thq}.  The Distance Conjecture is motivated by stringy examples.  In the string theory cases the tower of light states are either the excitations of a weakly-coupled, perturbative string, or a KK tower associated to a decompactification limit \cite{Lee:2019oct}.  In case the light tower of states corresponds to a critical string compactified to $d$ dimensions the relevant part of the effective action in string frame is given by 
\begin{equation}
    S \supset \int d^d x \sqrt{-g_S} \, \frac{M_s^{d-2}}{2(2\pi)^{d/2-1}} \,e^{-2\Phi} \left(R_S + 4 (\partial \Phi)^2\right) \,, 
\end{equation}
where $\Phi$ is the $d$-dimensional dilaton (including the contribution from the internal volume), $M_s$ the string scale and $g_S$ the metric in string frame. Note that this structure for the action is independent of the internal string geometry at the tree level due to the splitting of the CFT's associated to internal and macroscopic degrees of freedom.  Defining $\phi=\Phi-\Phi_0$, where $e^{\Phi_0}$ is the vev of the dilaton, and rescaling $g_E = e^{-\frac{4}{d-2} \phi} g_S$ one obtains the Einstein-frame action
$$S\supset \int \sqrt{-g_E}\ \frac{M_{\rm pl}^{d-2}}{2}\left[ R_E-\frac{4}{d-2}(\partial \phi_d)^2\right]\,,\qquad M_{\rm pl}^{d-2}= \frac{ e^{-2\Phi_0}}{(2\pi)^{d/2-1}} M_s^{d-2}\,.$$
In the following we always work in the Einstein frame and hence drop the index on the metric and the curvature. The canonically normalized field $\phi$ appearing in \eqref{distanceconj} is related to $\phi_d$ as $\phi=\frac{2}{\sqrt{d-2}}\phi_d$ such that we can read-off the parameter $\alpha$ appearing in \eqref{distanceconj} to be
$$\alpha_{\rm string} ={1\over \sqrt{d-2}}\,.$$
Consider now the case of a KK tower associated to the decompactification of a $d$-dimensional theory to a $D$-dimensional theory. Assuming that the internal $(D-d)$-dimensional manifold has radius $e^{\sigma} \equiv M_{{\rm pl},D} R$ we obtain the effective $d$-dimensional action in Einstein frame
\begin{equation*}
    S \supset \int d^dx \sqrt{-g}\ \frac{M_{\rm pl,d}^{d-2}}{2} \left[R - \frac{(D-2)(D-d)}{d-2}(\partial \sigma)^2\right]\,, \qquad M_{{\rm pl},d}^{d-2} = M_{{\rm pl},D}^{d-2}\, e^{(D-d)\sigma}\,. 
\end{equation*}
Now the mass of the KK-tower associated to the compactification is given by 
\begin{equation}
    m_{\rm KK} = \exp\left(-\frac{D-2}{d-2}\, \sigma\right)M_{{\rm pl},d}\,,  
\end{equation}
such that, using the canonically normalized field 
$$\phi= \left(\frac{(D-2)(D-d)}{(d-2)}\right)^{1/2}\,\sigma\,,$$
the parameter $\alpha$ in \eqref{distanceconj} for the KK-tower is given by
\begin{equation}\label{alphaKK}
\alpha_{\rm KK}={\sqrt{(D-2)\over (d-2)(D-d)}}\,. 
\end{equation}
Note that $\alpha_{\rm KK}> \alpha_{s}$.
The emergent string conjecture \cite{Lee:2019oct} states that these are the only two types of towers allowed.  If so, this implies that quite generally $\alpha  \geq {1\over \sqrt{d-2}}$, and
$$m\leq {\rm exp}(-\phi/ {\sqrt{d-2}}) \quad {\rm as} \ \phi\rightarrow \infty $$
Additional arguments for this bound have been given in \cite{Etheredge:2022opl}.

\subsection{Asymptotic Behavior of the Species Scale}
Due to the emergence of light degrees of freedom, one expects 
\begin{equation}\label{speciesvsmass}
    \Lambda_s=M_{\rm pl} N^{-1/(d-2)}\sim m^a\,,
\end{equation}
as we approach the boundaries of moduli space. Here $a$ is some order one number, and $m$ denotes the mass of the light tower as in \eqref{distanceconj}.  Indeed for the two cases of string and KK tower, it is easy to find the behavior of $\Lambda_s$ as we approach the boundary of moduli space.
For the case of a light string tower, $\Lambda_s=M_s$, i.e., the string scale.  To see this note that in perturbative string theory the higher-derivative corrections to the effective action have suitable powers of $M_s$ in the denominator, and the string scale serves as a cutoff at which the Einstein action receives strong corrections.
Since $M_s\sim M_{\rm pl}\,{\rm exp}(-\phi/\sqrt{d-2})$ we learn that in the emergent string case

\begin{equation}\label{lambdastring}
    \Lambda_s^{\rm string}\sim M_{\rm pl}\, {\rm \exp}(-\phi /\sqrt{d-2})\,.
\end{equation}
For the case of the KK tower, where the $d$-dimensional theory becomes $D$-dimensional, the species scale is the $D$-dimensional Planck scale, $M_{\rm pl,D}$.  This can be seen by noticing that as we approach the boundary the effective cutoff in the action is the higher-dimensional quantum gravity cutoff.  Since
$\Lambda_s^{\rm KK}=M_{{\rm pl},D}\sim M_{\rm pl,d} \,m_{\rm KK}^{(D-d)/(D-2)}$ we learn that, using \eqref{alphaKK},
\begin{equation}\label{lambdadecom}
\Lambda_s^{\rm KK}\sim M_{\rm pl,d}\,{\rm \exp}\left(-\phi {\sqrt{(D-d)\over (d-2)(D-2)}} \right)\,.    
\end{equation}
Assuming the emergent string conjecture this suggests that in either case we have
$$\Lambda_s\geq M_{\rm pl,d} \,{\rm exp}(-\phi /\sqrt{d-2}).$$
This in particular implies that as we approach the boundaries of moduli space we expect %
$$|\Lambda_s'/\Lambda_s |^2 \leq {1\over d-2}.$$
Indeed, using EFT arguments and that the contribution of $\phi$ to the higher-derivative gravitational terms should not be larger than the contribution of the massive states, we will argue in the next section that
\begin{equation}\label{lambda_bound}
    |\Lambda_s'/\Lambda_s |^2 \leq \mathcal{O}(1)\,.
\end{equation}
In particular in deriving this bound we do not use any assumptions coming from string theory and moreover we argue this holds not only asymptotically, but at all points in the moduli space including its interior. Interestingly, this bound gets saturated only at the boundaries where $\Lambda_s$ dies off exponentially.

\section{Higher-derivative Expansion and Species Scale}\label{sec:boundderivation}
We now would like to argue for a universal lower bound on $\Lambda_s'/\Lambda_s$ from consistency of the higher-derivative expansion of an effective theory of gravity. Therefore consider a simple theory of $d$-dimensional Einstein gravity coupled to a single real scalar field $\phi$ described by the two-derivative action 
\begin{equation}
    S_{\rm 2-der.} =  \int d^dx \sqrt{-g} \left(\frac{M_{\rm pl}^{d-2}}{2} R - \frac{1}{2} (\partial \phi)^2\right)\,. 
\end{equation}
The discussion below easily generalizes to many scalar fields, but for simplicity of notation we restrict here to the single field case.
In addition to these two-derivative terms, as we have already discussed, we generally expect corrections to Einstein gravity due to higher-dimension operators. We are particularly interested in gravitational higher-dimension operators involving the curvature $R$ as in \eqref{generalform} which are suppressed by suitable factors of $\Lambda_s$. As reviewed in the previous section in general the scale $\Lambda_s$ varies as we vary the scalar field $\phi$, cf. \cite{Long:2021jlv} and also \cite{vandeHeisteeg:2022btw} for a discussion of this scalar field dependence of the species scale in $\mathcal{N}=2$ theories in four dimensions. We now want to constrain this field dependence by requiring that the perturbative expansion in \eqref{generalform} is well-defined. From our discussion in the preceding section we know that the gravitational sector of the effective action \eqref{generalform} is valid up to the quantum gravity cutoff, $\Lambda_s$, due to the contribution of massive modes. More precisely by integrating out the massive modes the effective cutoff in the gravity sector can be lowered from $M_{\rm pl}$ to $\Lambda_s$. The idea we will pursue here is to divide the contribution of higher-energy modes to the effective action into two parts.  First the contribution of integrating out the massive modes with mass above $\Lambda_s$, and second the contribution of the short-distance modes of the other fields, in particular the massless fields, up to the species distance scale $\Lambda_s^{-1}$.
The basic idea is that the latter cannot dominate over contributions coming from the massive modes leading to the $\Lambda_s$ cutoff, since a single massless field alone cannot account for the entropy of the smallest black hole.  

\begin{figure}[!t]
\begin{center}
\includegraphics[width=0.66\textwidth]{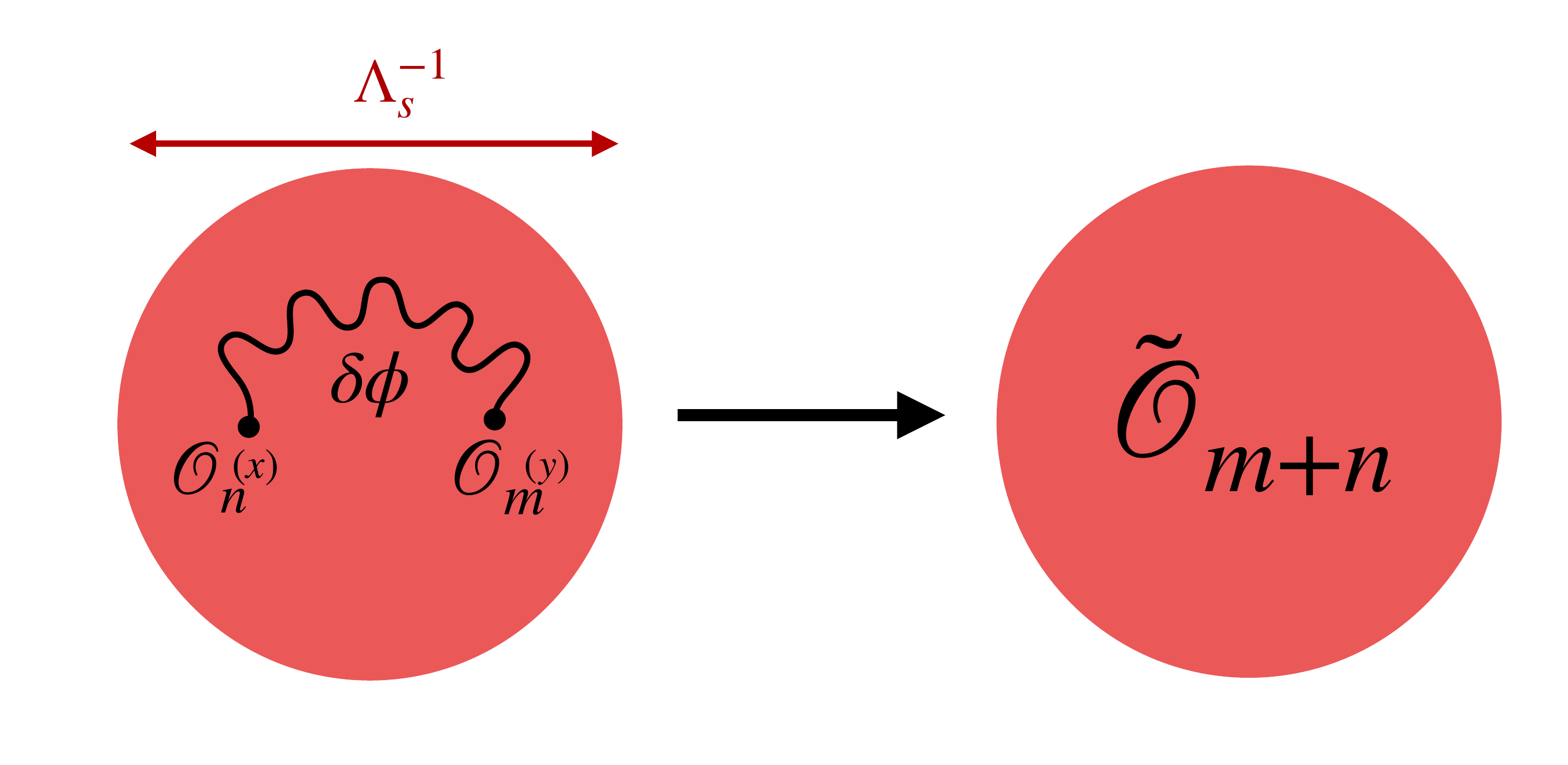}
\end{center}
\caption{An illustration of the generation of the higher-dimension operator $\tilde{\mathcal{O}}_{m+n}$ smeared over a ball of size $\Lambda_s^{-1}$ obtained by integrating out the small-distance modes of the scalar field exchange between two point-like operators $\mathcal{O}_n$ and $\mathcal{O}_m$.}\label{operatorsmearing}
\end{figure}

After the first step, namely the integrating out the contribution of massive modes above $\Lambda_s$, we expect the effective action to be of the form 
\begin{equation}
    S = \int \mathrm{d}^dx \, \sqrt{-g} \left[\frac{M_{\rm pl}^{d-2}}{2} \left(R +\sum_n  \,\frac{\mathcal{O}_n(R)}{\Lambda_s^{n-2}(\phi)}  \right) - \frac{1}{2}(\partial \phi)^2+\dots\, \right]. 
\end{equation}
In the second step we consider the effect of integrating out the UV modes of $\phi$ up to a distance scale $1/\Lambda_s$.
To this end let us first fix a vev $\langle \phi\rangle = \phi_0$ for the scalar field and consider small fluctuations $\delta \phi$ around this background, s.t. $\phi=\phi_0+\delta \phi$. Expanding 
\begin{equation}
   \Lambda_{\rm s} (\phi_0+\delta \phi) = \Lambda_{s}(\phi_0) + \delta \phi \, \Lambda'_{\rm s}(\phi_0) + \mathcal{O}\left( \delta \phi^2\right)\,
\end{equation}
we can generate interactions between the scalar fluctuation $\delta \phi$ and the operators $\mathcal{O}_n(R)$ by replacing
\begin{equation*}
     \frac{\mathcal{O}_n(R)}{\Lambda_{s}(\phi)^{n-2}} \rightarrow  \frac{\mathcal{O}_n(R) }{\Lambda_{s}(\phi_0)^{n-2}} \, - \frac{(n-2)\Lambda'_{s}(\phi_0)}{\Lambda_{s}(\phi_0)^{n-1}}\, \delta \phi\, \mathcal{O}_n(R)\,. 
\end{equation*}
We can now consider the scalar exchange between two operators $\mathcal{O}_n(R)$ and $\mathcal{O}_m(R)$. Given the above effective action this leads to the generation of an effective term
\begin{equation}\label{amplitude}
\sim \int \mathrm{d}^dx\mathrm{d}^dy \ M_{\rm pl}^{2(d-2)} \mathcal{O}_n(x)\frac{\Lambda'_{s}(\phi_0)}{\Lambda_{s}(\phi_0)^{n-1}} G_{\phi}(x,y)\mathcal{O}_m(y)   \frac{\Lambda'_{s}(\phi_0)}{\Lambda(\phi_0)_{s}^{m-1}}\,,
\end{equation}
where $G_{\phi}(x,y)\sim |x-y|^{2-d}$ is the Greens function for the scalar field. We now want to integrate out the high-energy modes of $\delta \phi$ with momenta $|p|$ between $M_{\rm pl}$ and $ \Lambda_{s}(\phi_0)$. This corresponds to performing one of the integrals in \eqref{amplitude} over a spacetime region defined by $ |x-y| \leq  \Lambda_{s}(\phi_0)^{-1}$.   In particular consider a smeared version of the gravitational operators
$$\tilde{\mathcal{O}}_k(R)(x)=\Lambda_s^d \int_{|y-x|\leq \Lambda_s^{-1}} d^dy \, \mathcal{O}_k(R)(y)\,,$$
which is suitable when we have a distance cutoff of $\Lambda_s^{-1}$.
Upon integrating out the UV modes of the $\delta \phi$ field with $|p|>\Lambda_s(\phi_0)$, the interaction in \eqref{amplitude} effectively generates a point-like interaction corresponding to a higher-dimension operator of dimension $m+n$, $\tilde{\mathcal{O}}_{m+n}(R)$, smeared over a ball of radius $\Lambda_s^{-1}$: 
\begin{equation}\label{Lambdasexpansion}
    \int d^d x\, \tilde{\mathcal{O}}_{m+n}(R)(x) \frac{M_{\rm pl}^{2d-4}(\Lambda'_{s}(\phi_0))^2}{\Lambda_{s}(\phi_0)^{n+m}}\,.
\end{equation}
The generation of this smeared operator is illustrated in Figure~\ref{operatorsmearing}. 
Therefore, we expect a correction to the effective action given by 
\begin{equation}\label{additionaloperator}
    S \supset M_{\rm pl}^{d-2} \int d^dx\,\sqrt{-g}\, M_{\rm pl}^{d-2}\frac{\Lambda'_{s}(\phi_0)^2}{\Lambda_s(\phi_0)^2}\frac{1}{\Lambda_{s}(\phi_0)^{n+m-2}} \tilde{\mathcal{O}}_{m+n}(R)\,,
\end{equation}
where we factored out $M_{\rm pl}^{d-2}$ to have the same normalization as in \eqref{generalform}. For the expansion in \eqref{generalform} to be under control we need to require that the term in the effective action corresponding to the $(m+n)$-dimensional operator generated by the scalar exchange of a single field is smaller than the one generated by integrating out the massive modes above $\Lambda_s$. In other words, the prefactor in \eqref{additionaloperator} needs to be smaller than $\Lambda_{s}(\phi_0)^{-m-n+2}$ leading to the constraint
\begin{equation}\label{QGbound}
    \left|\frac{\Lambda'_{s}(\phi_0)}{\Lambda_{s}(\phi_0)}\right|^2 \leq \frac{c}{M_{\rm pl}^{d-2}}\,,
\end{equation}
for some universal constant $c\sim \mathcal{O}(1)$ (which may depend on $d$). This implies that the quantum gravity cutoff cannot decrease/grow any faster than exponentially in $\phi$ with the precise exponent being dictated by $c$. The above discussion can easily be extended to the case of multiple scalar field $\phi_i$ in which case \eqref{QGbound} becomes 
\begin{equation}
    {\Bigg|\frac{\nabla \Lambda_s(\phi_i)}{\Lambda_s(\phi_i)}\Bigg|}^2 \leq \frac{c}{M_{\rm pl}^{d-2}}\,,
\end{equation}
with $$|\nabla \Lambda_s(\phi_i)|^2=\left(g^{ij} \partial_{\phi_i} \Lambda_s \partial_{\phi_j} \Lambda_s\right)\,, $$
where $g^{ij}$ is the inverse metric of the $\phi_i$ field space.\footnote{The argument that the contribution of massless modes to the entropy should be small may be questioned in cases where the number of massless modes $N_0$ is large.  But it was seen in \cite{Long:2021jlv} that even then, $\Lambda_s^{-(d-2)}>N_0$, and so the massless states do not make the dominant contribution to the black hole entropy at the species scale.  In the above we assume this is the case in general.  Without this assumption the above inequality would get a factor of $N_0$ on the right hand side, which in turn is also believed to be finite and depend only on $d$ which can thus be absorbed in the constant of order 1 in the bound.}
Notice that this bound is expected to hold not only asymptotically but at {\it all} points in the moduli\footnote{Note also that using the fact that $\Lambda_s <M_{\rm pl}$ we get the weaker inequality $|\Lambda_s'|^2\leq c/M_{\rm pl}^{d-4}$.} 
 
 It would be tempting to conjecture that in the above inequality $c={1/({d-2})}$, based on the fact that asymptotically the emergent string conjecture implies this bound.  However, as we will show in the next section in explicit examples $|\Lambda'/\Lambda|$ approaches the asymptotic value from above, and so the slope of $\Lambda_s$ at interior points would violate the bound \eqref{QGbound} for this value of $c$.  We thus see that $c$, which is an $\mathcal{O}(1)$ number that may depend on the dimension $d$, is strictly larger than ${1\over {(d-2)}}$.

Interestingly, as reviewed in the previous section, in asymptotic regions of the scalar field space the Distance Conjecture \cite{Ooguri:2006in} predicts the existence of a tower of states that become light exponentially in the field space distance, i.e. 
\begin{equation}\label{towerdecay}
    \frac{m_{\rm tower}}{M_{\rm pl}} \lesssim e^{-\alpha \phi/M_{\rm pl}^{(d-2)/2} }\,,
\end{equation}
for some $\alpha\sim \mathcal{O}(1)$. However, there is a priori no argument why the mass of the tower cannot decay faster than exponential even though all string theory examples indeed feature the behavior in \eqref{towerdecay}. As reviewed in the previous section, in asymptotic regions the species scale and the mass scale $m_{\rm tower}$ satisfy
\begin{equation}\label{speciesscaledecay}
    \frac{\Lambda_s}{M_{\rm pl}} \sim \left(\frac{m_{\rm tower}}{M_{\rm pl}}\right)^a\,,
\end{equation}
where the coefficient $a$ depends on the characteristics of the tower of light states. Now \eqref{QGbound} tells us that $\Lambda_s'/\Lambda_s$ is bounded from above. This, however, implies that for large $\phi$ also the mass of the light tower cannot decay faster than exponentially. Therefore the exponential decay predicted by the Distance Conjecture is in fact marginally allowed within a consistent expansion of higher-derivative corrections to Einstein gravity consistent with the black hole interpretation of the species scale.

\section{Species Scale Bound in Type II Compactifications}\label{sec:typeII}
In this section we are specializing to the case of Type II compactifications on Calabi--Yau threefolds leading to effective theories of gravity with $\mathcal{N}=2$ supersymmetry in four dimensions. For these theories the behavior of the species scale as a function of moduli has been discussed in \cite{vandeHeisteeg:2022btw} and (at least in the vector multiplet sector) can be computed explicitly from topological string theory. In section~\ref{sec:generalities} we first discuss the general properties of the species scale in the vicinity of singular points in the vector multiplet moduli space of Calabi--Yau compactifications of Type II string theory and then illustrate our findings in explicit examples in section~\ref{sec:examples}. 

\subsection{General Discussion}\label{sec:generalities} 
As argued in \cite{vandeHeisteeg:2022btw} for Type II compactifications on Calabi--Yau threefolds, the species scale can be related to the one-loop topological free energy $F_1$ which, from the perspective of the underlying 2d $\mathcal{N}=(2,2)$ CFT, is defined by the index \cite{Cecotti:1992vy,Bershadsky:1993ta}
\begin{equation}\label{F1def}
    F_1 = \frac12 \int_\mathcal{F} \frac{d^2\tau}{\tau_2} \Tr\left((-1)^F F_LF_R q^{H_0} \bar{q}^{\bar{H}_0}\right)\,,
\end{equation}
with $F_{L(R)}$ the left-(right-)moving fermion number, $\mathcal{F}$ the fundamental domain of $SL(2,\mathbb{Z})$ and $H_0$ the Hamiltonian of the CFT. Following \cite{vandeHeisteeg:2022btw} the species scale is related to $F_1$ via 
\begin{equation}\label{LambdasF1}
    \Lambda_s = \frac{M_{\rm pl}}{\sqrt{F_1}}\,,
\end{equation}
see also \cite{Cribiori:2022nke} for a discussion of the black hole perspective on this relation. In light of the discussion in the previous sections, this relation can most easily be seen by noticing that in the effective 4d $\mathcal{N}=2$ supergravity action $F_1$ appears as the coefficient of the higher-derivative term \cite{Bershadsky:1993cx,Antoniadis:1993ze}
\begin{equation}\label{effWsquared}
    S_{\rm eff}\supset \int \mathrm{d}^4 x\int  \mathrm{d}^4 \theta\,  F_1 \mathcal{W}^2\,,
\end{equation}
where $\theta$ are the fermionic superspace coordinates and $\mathcal{W}_{\mu \nu}$ is the Weyl superfield for which the component expansion reads \cite{DEROO1980175}
\begin{equation}
  \mathcal{W}_{\mu \nu} = T^-_{\mu \nu} - R^-_{\mu\nu\lambda\rho} \theta \sigma^{\lambda \rho} \theta+ \dots\,,
\end{equation}
where $T^-,R^-$ correspond respectively to the anti-self-dual components of the graviphoton field strength and the curvature. Finally the square of the Weyl superfield appearing in \eqref{effWsquared} is defined as 
\begin{equation}
    \mathcal{W}^2 =\mathcal{W}_{\mu \nu} \mathcal{W}^{\mu \nu} \,.
\end{equation}
Performing the superspace integral then yields\footnote{Notice that, as in \cite{Antoniadis:1993ze}, we are working in Lorentzian signature. In Euclidean signature the Weyl superfield is self-dual in its Lorentz indices which yields an $R_+^2$-term upon superspace integration (as e.g.~in \cite{Gopakumar:1998ii}).}
\begin{equation}\label{actionR2}
    S_{\rm eff} \supset -\int\mathrm{d}^4 x \, F_1 (R^-)^2 + \dots \,. 
\end{equation}
Comparing with the general expectation for the higher-derivative expansion \eqref{generalform} would then directly lead to \eqref{LambdasF1}.  However this point is a bit more subtle:  We need to include the contribution of states with mass larger than $\Lambda_s$, but the above term is the contribution of all massive fields in the EFT including those below $\Lambda_s$.  Still, the topological nature of the above term is responsible for its not depending on how many additional states we keep.  In particular due to supersymmetry no massive particle contributes to $F_1$, and it only receives contribution from BPS instantons.\footnote{This is valid as long as we avoid points at which BPS states become massless compared to the string scale, such as the conifold point. Note that this does not exclude singularities such as the large volume point, since in the perturbative string regime, i.e. $g_s< 1$, D0-branes remain massive compared to the string scale, but become massless compared to the 4d Planck scale. }   So at least in this regime we can trust the relation \eqref{LambdasF1}, as the states which would or would not be included in the EFT as we move in moduli space will not affect $F_1$.

By integrating the holomorphic anomaly equation a closed form for $F_1$ can be obtained \cite{Bershadsky:1993ta} which only depends on the vector multiplet moduli of the effective $\cN=2$ theory. In the following we concentrate on the vector multiplet sector of Calabi--Yau compactification of Type IIA for which $F_1$ reads
\begin{equation}\label{eq:defF1}
 F_1 = \frac12 \left(3+h^{1,1}+ \frac{\chi}{12}\right)K -\frac12 \log \det G_{i\bar j} + \log |f|^2\,. 
\end{equation} 
Here $h^{1,1}$ is the dimension of the vector multiplet moduli space and $\chi$ the Euler characteristic of the Calabi--Yau threefold. Furthermore, $K$ and $G_{i\bar j}$ are respectively the K\"ahler potential and metric on the vector multiplet moduli space, and $f$ is a holomorphic function that can be fixed by matching the asymptotics of $F_1$ at the boundaries of the moduli space.

Notice that since \eqref{F1def} diverges for the zero modes, whose contribution hence needs to be subtracted from $F_1$, the form for $F_1$ obtained by integrating the holomorphic anomaly equation is only defined up to an additive constant. This additive constant depends on the chosen background on which the one-loop topological string amplitude is computed. Therefore this constant can also depend on the background values for the scalars in the hypermultiplet sector. Let us stress, however, that expanding $\Lambda_s=M_{\rm pl}/\sqrt{F_1}$ as in \eqref{Lambdasexpansion} will only generate an interaction between $(R^-)^2$ and the scalars in the vector multiplet sector whereas an interaction between the hypermultiplet scalars and $(R^-)^2$ is forbidden by supersymmetry. 

Given the explicit form for $F_1$ --- and hence for $\Lambda_s$ --- we can now test the validity of our bound on $\Lambda_s'/\Lambda_s$ in the vector multiplet moduli space of type II Calabi--Yau compactifications. Before we consider explicit examples in the next section, let us first study the general behavior of $\Lambda_s'/\Lambda_s$ as we approach certain classes of boundaries of the moduli space. 

\subsubsection*{Large volume limit} 
We begin by considering limits towards the large volume point. In Type IIA on Calabi--Yau threefolds, such limits correspond to a one-dimensional decompactifications to M-theory signalled by a tower of light D0-brane states \cite{Grimm:2018ohb,Lee:2019oct}.\footnote{In order not to move in the hypermultiplet sector we need to keep the Type IIA 4d dilaton $e^{-2\phi_4} = g_s^{-2} \mathcal{V}_{\rm CY}$ constant. In the large volume limit this requires to co-scale $g_s^2 \sim \mathcal{V}_{\rm CY}$ leading to the D0-brane tower being the lightest tower of states.} For the gradient of the species scale this means we expect to find the asymptotic ratio $|\nabla \Lambda_s|/\Lambda_s \to 1/\sqrt{6}$ in the large volume limit. In the following we reproduce this scaling from $F_1$, and while also paying special attention to the corrections to this behavior. 

Let us begin by parametrizing the large volume limit in terms of the moduli fields $t^i = a^i+i v^i$, where $a^i$ are the axions with a shift symmetry $a^i \to a^i+1$ and we take the large field limit for the saxions $v^i \to \infty$. We choose to scale all saxion fields homogeneously with some parameter $s $, i.e., we consider the scaling
\begin{equation}
    v^i = s \, \hat{v}^i\, ,
\end{equation}
where we send $s \to \infty$ and the $\hat{v}^i$ are kept constant.  To leading order the topological genus-one free energy behaves as 
\begin{equation}\label{eq:F1LCSlead}
    F_1 \to \frac{ c_{2,i}\hat{v}^i}{12} 2\pi  s \, ,
\end{equation}
where $c_{2,i}$ denote the integrated second Chern class of the Calabi-Yau threefold. Since we are interested in the corrections to the leading behavior of $|\nabla \Lambda_s|/\Lambda_s $, let us also compute these corrections to $F_1$ itself, which turn out to be proportional to $\log s$. Such terms originate from the K\"ahler potential $K$ and the K\"ahler metric $\log \det G_{i\bar{j}}$ in \eqref{eq:defF1}, which behave respectively as
\begin{equation}
    K \supset -3\log[s]\, , \qquad \log \det G_{i\bar{j}} \supset -2h^{1,1} \log[s]\, .
\end{equation}
Here we used that the K\"ahler potential is given by to logarithm of the volume of the Calabi--Yau which scales as $s^3$, and each of the $h^{1,1}$ eigenvalues of the K\"ahler metric scales as $s^{-2}$. 
Collecting their individual contributions we find the asymptotic behavior of $F_1$, including the first sub-leading term, to be given by
\begin{equation}\label{eq:F1LCS}
    F_1 \to \frac{ c_{2,i}\hat{v}^i}{12} 2\pi s -\frac{1}{4}(18+h^{1,1}+h^{2,1})\log[s]+\mathcal{O}(s^0)\, ,
\end{equation}
where we used that $\chi = 2(h^{2,1}-h^{1,1})$. We next compute the gradient of the species scale corresponding to $F_1$ using \eqref{LambdasF1}. Given the metric $G_{ss} = 3/(2s)^2$ and the behavior stated above we find
\begin{equation}
\begin{aligned}
    \frac{|\nabla \Lambda_s|}{\Lambda_s} &= \frac{1}{2}\sqrt{G^{ss}/2} \frac{|\partial_s F_1|}{F_1} \\
    &= \frac{1}{\sqrt{6}} + \frac{\sqrt{3}}{2\sqrt{2}\pi s (c_{2,i}\hat{v}^i)} \left[(18+h^{1,1}+h^{2,1})\log[s] +\mathcal{O}(s^0)\right]+\mathcal{O}(s^{-2})\, .
\end{aligned}
\end{equation}
Since the leading correction is strictly positive this tells us that the gradient of the species scale approaches the asymptotic value $1/\sqrt{6}$ from above. In particular, this means that $|\nabla \Lambda_s|/\Lambda_s$ will reach a maximum between the desert point and the large volume point where it takes values larger than $1/\sqrt{6}$. This point is illustrated by figure \ref{fig:X5slice} where we provided a plot for the quintic near the large volume point.

\subsubsection*{Decompactification limits to 6d}
The second class of infinite distance limits for which we can study the asymptotic behavior for $F_1$ corresponds to decompactification limits to 6d. These arise if the Calabi--Yau threefold allows for an elliptic/genus-one fibration $X_3: T^2\rightarrow B_2$ \cite{Lee:2019oct}. In this case we can split the K\"ahler moduli into two sets $(t^a_f, t^\alpha_b)= (b^a_f + iv^a_f, b^\alpha_b + iv^\alpha_b)$ depending on whether they parametrize the volume of fibral or base curves, respectively. We can now consider the limit 
\begin{equation}
    v^a_f = \hat{v}^a_f\,,\quad v^\alpha_b = s \, \hat{v}^\alpha_b\,,
\end{equation}
where we send $s\rightarrow \infty$. As described in \cite{Lee:2019oct} this limit (accompanied by a suitable co-scaling of $g_s$) corresponds to a decompactification to F-theory on $X_3$. In the asymptotic region, the K\"ahler potential is given by 
\begin{equation}\label{eq:Kpot6d}
    K = - \log\left[\frac12 \left(\sum_a c_a v^a_f\right) \eta_{\alpha \beta} v^\alpha_b v^\beta_b+ \mathcal{O}(s) \right]\,,
\end{equation}
for some coefficients $c_a$ and $\eta$ the intersection form on $B_2$, with $\mathcal{O}(s)$ representing subleading terms that are at most linear in the base moduli $v_b^\alpha$. In order to deal with the metric determinant term $\log \det G_{i \bar j}$ in $F_1$ it proves to be useful to rewrite
\begin{equation}\label{F1smallg}
    F_1 = \frac12 \left(3 + \frac{\chi}{12}\right) K - \frac12 \log \det g + \log |f|^2\,,
\end{equation}
where we rescaled the metric by $g= e^{-K} G$ such that 
\begin{equation}
    g= \frac{3}{2} \left(\frac{3}{2} \frac{\mathcal{V}_i \mathcal{V}_j}{\mathcal{V}} - \mathcal{V}_{ij}\right)\,,
\end{equation}
where $\mathcal{V} = \kappa_{ijk} v^i v^j v^k$, $\mathcal{V}_i = \kappa_{ijk} v^j v^k$, and $\mathcal{V}_{ij} = \kappa_{ijk} v^k$ for $\kappa_{ijk}$ the triple-intersection numbers on $X_3$. Specializing now to the intersection numbers in \eqref{eq:Kpot6d} for the elliptic fibration, and separating metric components for the fibral and base moduli $t^a_f$ and $t^\alpha_b$, in the limit $s\rightarrow \infty$ we find
\begin{equation}
  g_{ab} = g_{a b}^{(2)}\, s^2 + g_{a b}^{(1)} \,s + \mathcal{O}( s^0)\,,\qquad g_{\alpha \beta} = g_{\alpha \beta}^{(0)} + \mathcal{O}(1/s)\,, \quad g_{a\alpha} = g_{a \alpha}^{(1)} \,s + \mathcal{O}(s^0)\,.
\end{equation} 
The scaling of $\log \det g$ now depends on the rank of the constant matrices $g_{ab}^{(2)}$, $g_{ab}^{(1)}$, $g_{\alpha \beta}^{(0)}$, and $g_{a \alpha}^{(1)}$. Borrowing the results of \cite{Cota:2022maf} we find that, except for $g_{ab}^{(2)}$ which has rank 1, these constant matrices have full rank. Using this we find the asymptotic behavior of $F_1$ to be given by 
\begin{equation}\label{F16decomp}
    F_1 \simeq \frac{2\pi }{12} c_{2,\alpha} \hat{v}_b^{\alpha} s - \frac{1}{12}\left(42+4 h^{1,1} + 2 h^{2,1} - 6 h^{1,1}(B_2)\right) \log[s]+\mathcal{O}(s^0)\,,
\end{equation}
where $h^{1,1}(B_2)$ denotes the number of K\"ahler moduli of the base $B_2$. We can use this asymptotic behavior to compute the slope of $\Lambda_s$ to be 
\begin{equation}
    \frac{|\nabla \Lambda_s|}{\Lambda_s} = \frac{1}{2} \sqrt{G^{ss}/2} \frac{|\partial_{s} F_1|}{F_1} = \frac{1}{2} + \frac{1}{2\pi (c_{2, \alpha} \hat{v}^\alpha_b) s}[(42+4h^{1,1}+2h^{2,1}-6h^{1,1}_{\rm exc})\log[s]+\mathcal{O}(s^0)]\, ,
\end{equation}
where we used $G_{ss}=1/2s^2$. The leading term $1/2$ matches with the expectation 
$$\sqrt{\frac{(D-d)}{(d-2)(D-2)}}=\frac12\,,$$
for a decompactification from $d=4$ to $D=6$ dimensions.

\subsubsection*{Emergent string limits}
One can repeat the analysis for the sub-leading terms in $F_1$ in the vicinity of points corresponding to emergent string limits. In these limits the leading tower of states arises from excitations of a string obtained by wrapping an NS5-brane on a K3 sub-manifold in the internal Calabi--Yau threefold. In this case, the species scale is simply set by the tension of this string.  

In one-dimensional moduli spaces we can realize emergent string limits at K-points in the moduli space, as we will discuss in section \ref{sec:examples} for an explicit example. To illustrate the features of the emergent string limits it is best to work instead with multi-moduli Calabi--Yau threefolds that allow for a K3-fibration. Given the fibration structure it is natural to split the moduli as $t^i=(t^1,t^K)$ where $t^1=b^1+iv^1$ is the complexified volume of the $\mathbb{P}^1$ base and $t^K=b^K + iv^K$ are complexified volumes of curves in K3-fibers. The emergent string limit now corresponds to 
\begin{equation}\label{Klimit}
    v^1 = s\, \hat{v}^1 \,,\qquad v^K = \hat{v}^K\,,\quad s\rightarrow \infty
\end{equation}
where $\hat{v}^1$ and $\hat{v}^K$ are kept constant in the limit.\footnote{Similarly to the large volume limit we need to co-scale the Type IIA string coupling $g_s^2 \sim v^1$ in order to keep constant the 4d dilaton, see \cite{Lee:2019oct} for a careful treatment of this co-scaling.} As we are interested in the subleading behavior of $|\nabla \Lambda_s|/\Lambda_s$, let us also compute corrections proportional to $\log s$. Again such terms come from the K\"ahler potential $K$ and the K\"ahler metric $\log \det g$ in \eqref{F1smallg}.  To evaluate the latter we need to know the scaling of the entries of $g_{i\bar{j}}$. Given the split of the moduli into $(v^1, v^K)$ and the limit \eqref{Klimit} the scalings are 
\begin{equation}
    g_{11}= g_{11}^{(0)}  s^{-1} + \mathcal{O}\left(s^{-2}\right)\,,\qquad  g_{KL} = g_{KL}^{(0)}s  + \mathcal{O}\left(s^0\right) \,,\quad g_{1K}=\mathcal{O}\left(s^0\right)\,. 
\end{equation}
for some constant function $g_{11}^{(0)}$ and constant matrix $g_{KL}^{(0)}$. The scaling of $\log \det g$ now crucially depends on the rank of $g_{KL}^{(0)}$. For K3-fibered Calabi--Yau threefolds the rank of this matrix can be calculated following the analysis of \cite{Cota:2022maf}. Therefore we should split the moduli $v^K$ into two different sets depending on whether the corresponding curve is located in a generic, irreducible K3-fiber or located in an exceptional, reducible K3-fiber. In general the K3-fiber can degenerate into multiple components over multiple points in the base. If $h^{1,1}_{\rm exc}$ is the number of independent, exceptional fiber components arising over these degenerations, there are in total $h^{1,1}_{\rm exc}+1$ fibral divisors in the K3-fibration, accounting also for the class of the generic fiber. In addition there are $h^{1,1}_{\rm gen}$ divisors associated to the sections of the fibration such that
\begin{equation}
    h^{1,1}(X_3) = 1 + h^{1,1}_{\rm gen} + h^{1,1}_{\rm exc}\,. 
\end{equation}
The results of \cite{Cota:2022maf} now imply that in case $h^{1,1}_{\rm exc}\neq 0$ the matrix $g_{KL}^{(0)}$ does not have full rank but its rank is given by  
\begin{equation}
    \text{rk}\left(g^{(0)}_{KL}\right) = h^{1,1}_{\rm gen}\,. 
\end{equation}
In contrast the subleading, constant term in $g_{KL}$ has full rank such that we find 
\begin{equation}
    \log \det g \sim (h^{1,1}(X_3) - h^{1,1}_{\rm exc} - 2) \log[s]\,, 
\end{equation}
where we also took into account the scaling of $g_{11}$. In total the asymptotic behavior of $F_1$, including the first sub-leading term, is hence given by 
\begin{equation}\label{eq:F1Kpoint}
    F_1 \to 4\pi s -\frac{1}{12}(6+5h^{1,1}+h^{2,1}-6h^{1,1}_{\rm exc})\log[s]+\mathcal{O}(s^0)\, ,
\end{equation}
where we used that for K3-fibrations $c_{2,1}=24$. Using this result we can compute the gradient of the species scale. For large $s$ we find the behavior
\begin{equation}
    \frac{|\nabla \Lambda_s|}{\Lambda_s} = \frac{1}{2} \sqrt{G^{ss}/2} \frac{|\partial_{s} F_1|}{F_1} = \frac{1}{\sqrt{2}} + \frac{1}{2\sqrt{2}\pi s}[(6+5h^{1,1}+h^{2,1}-6h^{1,1}_{\rm exc})\log[s]+\mathcal{O}(s^0)]\, ,
\end{equation}
where $G_{ss}=1/(2s)^2$, and we suppressed subleading $v^K$-dependent terms. The leading term $1/\sqrt{2}$ matches precisely with the expected value $1/\sqrt{d-2}$ for a 4d emergent string limit.

\subsubsection*{Conifold}
Let us now turn to the vicinity of a conifold point which, unlike the previous limits, is a point at finite distance. We can choose a local coordinate $\mu$ on the moduli space in which the conifold point is located at $\mu=0$. In the vicinity of this point $F_1$ behaves to leading order like \cite{Vafa:1995ta}
\begin{equation}\label{F1coni}
    F_1 = -\frac{1}{12} \log |\mu|^2 +\dots\,,
\end{equation}
whereas the K\"ahler potential is given by 
\begin{equation}\label{kahlerconi}
    K=-\log(|\mu|^2 \log |\mu|^2 + \text{const.})\,. 
\end{equation}
Using these expressions we find 
\begin{equation}
    \frac{|\nabla \Lambda_s|^2}{\Lambda_s^2} \sim \left| \frac{1}{|\mu|^2 \left(\log|\mu|^2\right)^3}\right|\,,\qquad \text{as}\quad \mu\rightarrow 0\,,
\end{equation}
which diverges as $|\mu|\rightarrow 0$. At first sight this seems to be in tension with our bound \eqref{QGbound} implying that the perturbative expansion of the effective action of the theory breaks down. This is, however, to be expected for $|\mu|\rightarrow 0$ due to the appearance of the additional light D-brane state at the conifold point which invalidates the effective action we started with. As we discussed below  \eqref{actionR2}, in our effective theory --- in which we defined $\Lambda_s$ via $F_1$ --- this BPS state is integrated out and we therefore expect that the higher-derivative expansion breaks down as we approach the conifold point. In this sense the divergence of $|\nabla \Lambda_s|/\Lambda_s$ is just another symptom of the break-down of the EFT in the vicinity of the conifold point. 

For a correct treatment of the conifold locus we have two options: on the one hand, we can integrate-in the conifold state to get a consistent theory of Einstein gravity which now has an additional light hypermultiplet. As described in the previous section, the higher-derivative term in \eqref{actionR2} is obtained by integrating out all states above the massless level, including the conifold state. Hence integrating-in the conifold state changes this higher-derivative term such that, as a consequence, $\Lambda_s$ is not anymore simply given by the topological genus-one free energy $F_1$. On the other hand, we can also approach the conifold point while staying in the conformal field theory picture by taking $\mu\rightarrow 0$ while keeping 
\begin{equation}
    \hat{\mu}=\frac{\mu}{g_s} =\text{const.}
\end{equation}
This corresponds to a co-scaling of the string coupling $g_s$ as we approach the conifold point which ensures that the mass of the light D-brane remains above the string scale. In this case, $\Lambda_s$ continues to be given by $F_1$ as in \eqref{F1coni}. As detailed in \cite{vandeHeisteeg:2022btw} the divergence of $F_1$ can now be interpreted as coming from KK-modes along a tube of length $\log |\mu|$ that develops in the internal geometry in the vicinity of the conifold in the conformal field theory picture. To describe the metric on the field space metric we need to replace $\mu\rightarrow \hat{\mu}$ in the K\"ahler potential \eqref{kahlerconi}. With this replacement we can evaluate
\begin{equation}
    \frac{|\nabla \Lambda_s|^2}{\Lambda_s^2} = g^{\mu \bar \mu} \frac{\partial_\mu  \Lambda_s \partial_{\bar \mu}\Lambda_s }{\Lambda_s^2}= \frac{g_s^2}{\log|\hat{\mu}|^2} \frac{1}{|\mu|^2\log|\mu|^2 }\,,
\end{equation}
where we used $g^{\mu \bar \mu} = g_s^2 g^{\hat \mu \hat{\bar \mu}}= g_s^2/ \log|\hat{\mu}|^2$. Notice that since $g_s$ resides in a hypermultiplet we do not need to include a term including the derivatives of $\Lambda_s=1/\sqrt{F_1}$ w.r.t. to $g_s$ since no interaction between the dilaton and $(R^-)^2$ is allowed by supersymmetry. Using $\hat{\mu}=\text{const.}$ we find 
\begin{equation}
    \frac{|\nabla \Lambda_s|^2}{\Lambda_s^2} \rightarrow 0\,\qquad \text{for} \qquad \mu\sim g_s \rightarrow 0\,,
\end{equation}
which parametrically satisfies our bound \eqref{QGbound}.

\subsubsection*{The correction to the slope of $\Lambda_s$}
To summarize, in the three classes of infinite distance limits that exist for Type II compactifications on Calabi--Yau threefolds we found that our bound \eqref{QGbound} parametrically saturated and that asymptotically $F_1$ behaves as 
\begin{equation}
    F_1 = \frac{2\pi}{12}c_2 s - \beta \log s +\mathcal{O}(s^0)\,.
\end{equation}
Here $c_2$ is the relevant part of the second Chern class of the Calabi--Yau threefold, and $\beta$ is the coefficient of the logarithmic correction which in the three classes of infinite distance limits is given by 
\begin{equation}\begin{aligned}\label{betas}
    \text{Large Volume:} \qquad \beta_{\rm LV} &\stackrel{\eqref{eq:F1LCS}}{=} \frac{1}{4}(18+h^{1,1}+h^{2,1}) \,,\\ 
    \text{6d Decompactification:} \qquad \beta_{6d} &\stackrel{\eqref{F16decomp}}{=} \frac{1}{12}\left(42+4 h^{1,1} + 2 h^{2,1} - 6 h^{1,1}(B_2)\right)\,, \\    
    \text{Emergent String:}\qquad \beta_{\rm ES} &\stackrel{\eqref{eq:F1Kpoint}}{=} \frac{1}{12}(6+5h^{1,1}+h^{2,1}-6h^{1,1}_{\rm exc}) \,.
\end{aligned}\end{equation}
Depending on whether $\beta$ is positive (negative) the slope $|\nabla \Lambda_s|/\Lambda_s$ approaches its asymptotic value from above (below). In the majority of the examples we tested we found this coefficient to be positive, though for the emergent string case there seem to exist geometries for which $\beta_{\rm ES}<0$. Let us elaborate on these findings for the different types of limits:
\begin{itemize}
    \item In the large volume limit $\beta_{\rm LV}$ is manifestly positive implying that in this class of limits the slope of $\Lambda_s$ approaches its asymptotic value from above.
    \item  For genus-one fibered Calabi--Yau threefolds $\beta_{6d}$ can in principle become negative for geometries with large $h^{1,1}(B_2)$.  Such geometries have been studied, e.g., in \cite{Morrison:2012js} where for $h^{1,1}(B_2)\gg 1$ a scaling  $h^{1,1}\sim \frac{5}{2} h^{1,1}(B_2)$ was found. This implies that even for large $h^{1,1}(B_2)$ the coefficient of the logarithmic correction to $F_1$ satisfies $\beta_{6d}>0$. Still, we are not aware of a general proof that $\beta_{6d}$ are positive for any genus-one fibered Calabi--Yau threefold. 
    \item For K3 fibrations (emergent string limits) we are not aware of a general bound for $h^{1,1}_{\rm exc}$ in terms of $h^{2,1}$ and $h^{1,1}$ (apart from the trivial $h^{1,1}_{\rm exc}<h^{1,1}$). Let us notice, however, that for Type IIA compactification with a \emph{perturbative} heterotic dual, i.e., in case there are no space-time filling NS5-branes in the heterotic theory, $h^{1,1}_{\rm exc}=0$, and again $\beta_{\rm ES}$ is manifestly positive. Calabi--Yau geometries with K3-fibrations that do not have a perturbative heterotic dual were for instance discussed in \cite{Braun:2016sks}. An example of a K3-fibered Calabi--Yau threefold that seemingly evades $\beta_{\rm ES}\geq 0$ is given by the mirror octic which, following the analysis of \cite{Doran_2016}, has $h^{2,1}=1$ and $h^{1,1}=149$, with the number of exceptional fiber components equal to $h^{1,1}_{\rm exc} = 129$.\footnote{The Calabi-Yau hypersurface in $\mathbb{P}_{1,1,1,1,4}[8]$ belongs to the set of 14 one-parameter hypergeometric variations of Hodge structure. In \cite{Doran_2015} it was found that their mirror Calabi-Yau manifolds all admit a K3-fibration, with the mirror $X_{1,1,1,1,4}[8]$ fibered by the mirror quartic. In \cite{Doran_2016} a detailed general analysis of the fiber components was performed for such K3 fibrations. In our case this yields a decomposition $h^{1,1}=1+19+129$, with the first two terms corresponding to the $\mathbb{P}^1$ base and the curves in the generic K3-fiber, while the remainder corresponds to the exceptional components.} These Hodge numbers then lead to $\beta_{\rm ES}=-11/6$ for the mirror octic. It would be important to confirm that, indeed, the slope $|\nabla \Lambda_s|/\Lambda_s$ approaches its asymptotic value from above by direct computation of the exact $F_1$ for the mirror octic. 
\end{itemize}
Given the slope of $\Lambda_s$ in the emergent string limits with $\beta_{\rm ES}>0$ (such as the limits with perturbative heterotic dual), the naive bound 
\begin{equation}
    \left|\frac{\nabla \Lambda_s}{\Lambda_s}\right|^2 \leq \frac{ M_{\rm pl}^{2-d}}{d-2}\,, 
\end{equation}
can be explicitly violated. This implies that the constant $c$ appearing in \eqref{QGbound} needs to be larger than $1/(d-2)$. Since $|\nabla \Lambda_s|=0$ at the desert point in the interior of the moduli space where $\Lambda_s$ is minimized, the slope $|\nabla \Lambda_s|/\Lambda_s$ needs to have a maximum if $\beta>0$ somewhere between the asymptotic and the desert point. To determine the constant $c$ in \eqref{QGbound} a closer analysis of this maximum is necessary, which --- as we will see explicitly for the example of $(K3\times T^2)/\mathbb{Z}_2$ in the next subsection --- is sensitive to the additive constant to $F_1$.

\subsection{Examples}\label{sec:examples}
We now turn to explicit examples of moduli spaces arising from type II string compactifications and analyse the profile of the (canonically normalized) gradient $|\nabla \Lambda_s|/\Lambda_s$ of the species scale $\Lambda_s$. Following the previous general discussion we identify the species scale with $F_1$ as
\begin{equation}\label{eq:deflambda}
    \Lambda_s = \frac{1}{\sqrt{F_1+N_{\rm des}}}\, ,
\end{equation}
where $N_{\rm des}$ accounts for the additive constant in the definition of $F_1$ that is not captured by integrating the holomorphic anomaly equation. In the following we normalize $F_1$ such that at the global minimum of $F_1$ we have $F_1=0$. In this normalization $N_{\rm des}$ indeed corresponds to the number of light states at the desert point.

\subsubsection*{Example 1: \texorpdfstring{$(K3\times T^2)/\mathbb{Z}_2$}{K3 x T2}}
We begin our investigation with the simple case where the target space of the topological string is $T^2$. From the perspective of Type IIA Calabi-Yau compactifications this can be viewed as studying the $T^2$-dependence of the species scale for a compactification on the Enriques Calabi--Yau $(\rm{K3}\times T^2)/\mathbb{Z}_2$, while we suppress all dependence on the K3-moduli. The genus-one topological free energy is given by \cite{Bershadsky:1993ta}
\begin{equation}\label{eq:F1T2}
  F_1 = -6\log \left[i(\bar{t}-t) |\eta^2(t)|^2 \right]  +6\log\left[\frac{3\Gamma(\frac{1}{3})^{6}}{16\pi^4}\right]\, ,
\end{equation}
where $t=a+is$ is the complexified K\"ahler modulus of the torus.\footnote{Compared to the original result of \cite{Bershadsky:1993ta} for $T^2$, for the Enriques Calabi--Yau an additional factor of six is required \cite{Grimm:2007tm}.} The additive constant appearing in \eqref{eq:F1T2} is such that $F_1(e^{2\pi i /3})=0$ vanishes at the desert point. This desert point was originally identified in \cite{Long:2021jlv} as the point where the tension of Type IIB $(p,q)$-strings is maximized; in \cite{vandeHeisteeg:2022btw} it was verified that $F_1$ is minimized at this point as well, thereby matching with the expected behavior of the number of species at the desert. Taking the large-volume limit for the $T^2$, i.e.~$s \to \infty$, corresponds to an emergent string limit: Enriques Calabi-Yau can be viewed as a K3-fibration, and in this limit NS5-branes wrapped on the K3 submanifold become tensionless. 

\begin{figure}[!t]
\begin{center}
\includegraphics[width=10cm]{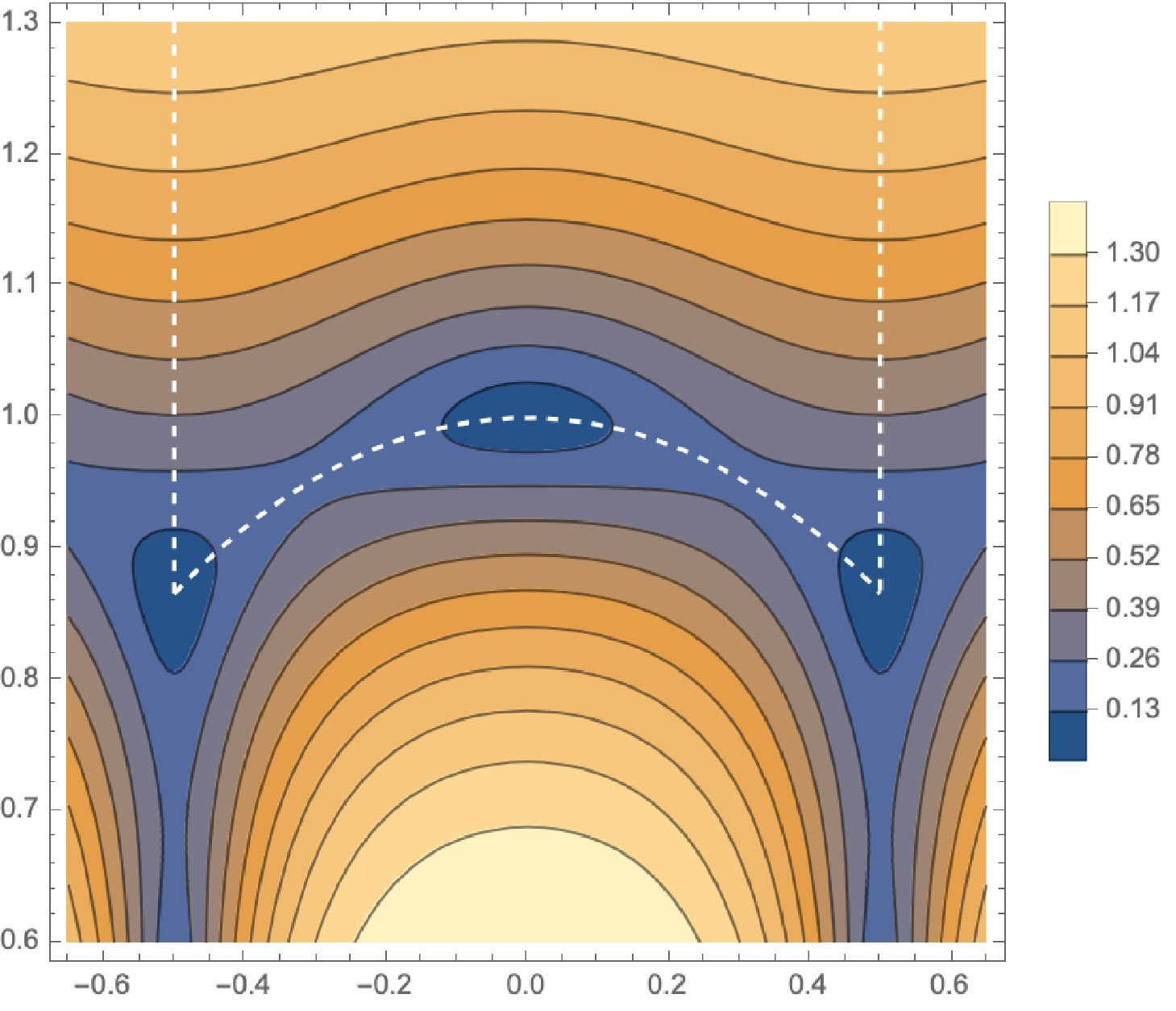}
\end{center}
\begin{picture}(0,0)\vspace*{-1.2cm}
\put(223,22){$a$}
\put(88,155){$s$}
\end{picture}\vspace*{-0.8cm}
\caption{\label{fig:T2contour} Contour plot of the gradient of the species scale $|\nabla \Lambda_s|/\Lambda_s$ in the upper half plane for $T^2$. The fundamental domain has been indicated by a dashed, white line. The zeroes of the gradient are located in the blue valleys, corresponding to the cusps $t=(\pm 1+i\sqrt{3})/2$ and $t=i$. The cusps $t=(\pm 1+i\sqrt{3})/2$ are global maxima of $\Lambda_s$, while $t=i$ is a local maximum.}
\end{figure}
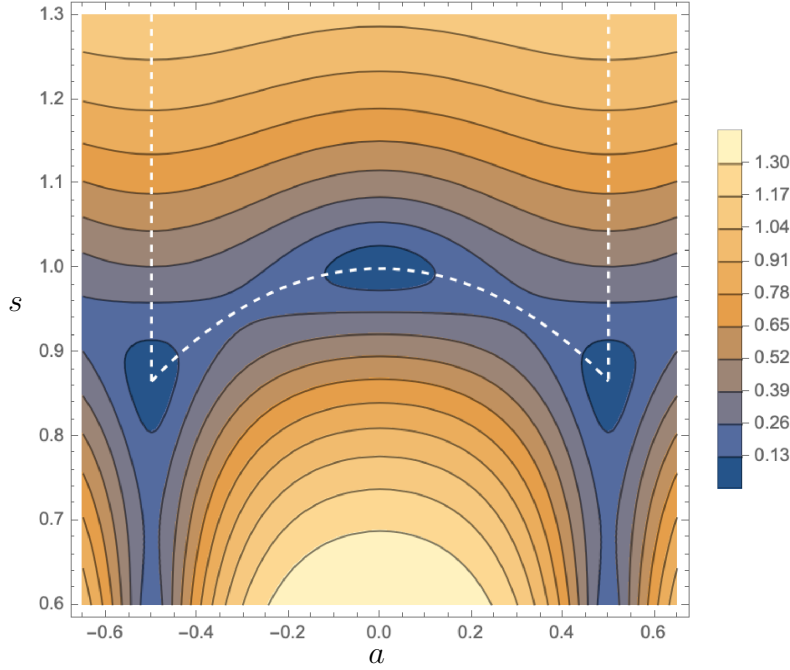

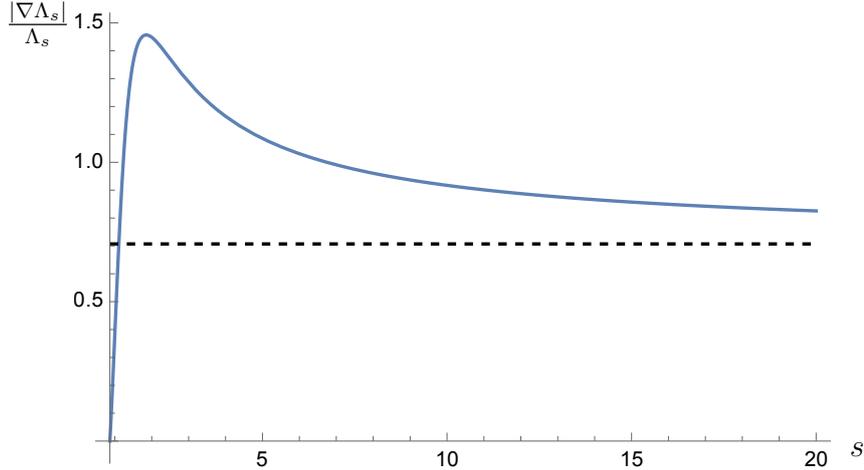
\begin{figure}[!ht]
\begin{center}
\includegraphics[width=10cm]{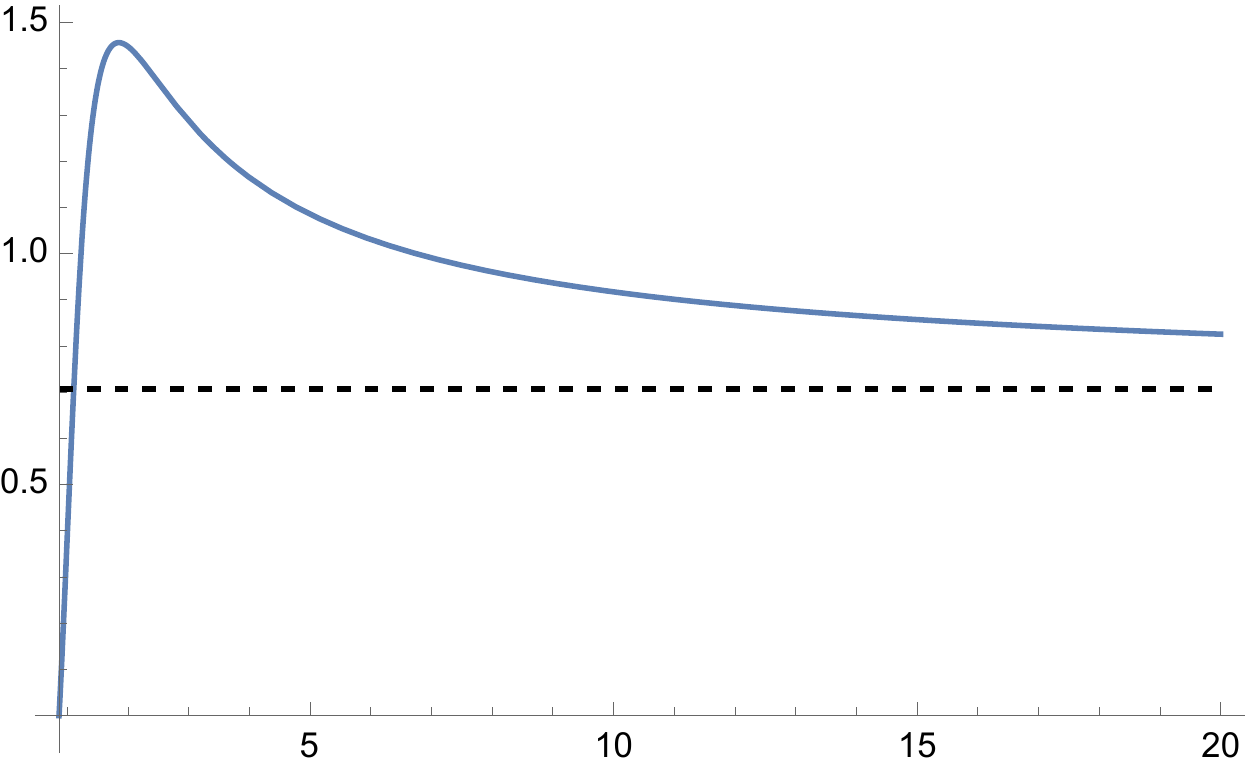}
\end{center}
\begin{picture}(0,0)\vspace*{-1.2cm}
\put(74,195){$\frac{|\nabla \Lambda_s|}{\Lambda_s}$}
\put(390,35){$s$}
\end{picture}\vspace*{-0.8cm}
\caption{\label{fig:T2slice} Plot of the gradient of the species scale $|\nabla \Lambda_s|/\Lambda_s$ in the upper half plane for $T^2$ along a fixed axion slice $a=-1/2$. The dashed black line indicates the bound $|\Lambda_s|/\Lambda_s = 1/\sqrt{2}$ towards which the contour asymptotes at large $s$. We have conservatively set the number of species at the desert to $N_{\rm des}=1$. Irrespective of the value of $N_{\rm des}$ we find that $|\nabla \Lambda_s|/\Lambda_s >1/\sqrt{2}$ has a maximum that exceeds the dimension-dependent bound $1/\sqrt{d-2}$ since $d=4$. By increasing $N_{\rm des}$ we can decrease this maximum and move it to larger values of $s$, as is described by equations \eqref{eq:T2max} and \eqref{eq:T2grad}.}
\end{figure}

In order to build intuition for the species scale we have provided figures of the gradient $|\nabla \Lambda_s|/\Lambda_s$: in figure \ref{fig:T2contour} we provide a contour plot over the upper-half plane; in figure \ref{fig:T2slice} we consider a slice with fixed axion $a = -\frac{1}{2}$, starting from the desert point $s = \sqrt{3}/2$ to large volume $s = \infty$. The former allows us to gain some general intuition for the maxima of $\Lambda_s$, both global (the desert point $t=e^{2\pi i/3}$) and local (the self-duality point $t=i$). The latter allows us to study the profile of $|\nabla \Lambda_s|/\Lambda_s$ more carefully, interpolating between the zero at the desert point and the exponential rate $|\nabla \Lambda_s|/\Lambda_s = 1/\sqrt{2}$ in the large volume limit.

Complementary to these figures, let us expand the slope $|\nabla \Lambda_s|/\Lambda_s$ in the scaling limit $s\gg 1$. Starting with $F_1$ given by \eqref{eq:F1T2}, we find in the limit $s\rightarrow \infty$
\begin{equation}\label{eq:F1T2asymp}
    F_1 = 2 \pi s - 6 \log[s]+6\log\left[\frac{3\Gamma(\tfrac{1}{3})^6}{32\pi^4}\right]+\mathcal{O}(e^{-2\pi s})\, .
\end{equation}
Notice that the coefficient of the $\log[s]$ term agrees with \eqref{eq:F1Kpoint} for the Hodge numbers $h^{2,1}=h^{1,1}=11$ of the Enriques Calabi-Yau ($h^{1,1}_{\rm exc}=0$ for $(K3\times T^2)/\mathbb{Z}_2$). Next we determine the behavior of the species scale gradient $|\nabla \Lambda_s|/\Lambda_s$. Taking \eqref{eq:deflambda} for the species scale and the hyperbolic metric $G_{t\bar{t}} = 1/(2s)^2$ we find that
\begin{equation}\label{eq:T2approx}
    \frac{|\nabla \Lambda_s|}{\Lambda_s} = \frac{1}{\sqrt{2}} + \frac{6}{2\sqrt{2}\pi s} \bigg(\log[s]-\frac{1}{6}N_{\rm des}+\log\left[\frac{3\Gamma(\tfrac{1}{3})^6}{32\pi^4}\right] \bigg)+\mathcal{O}(s^{-2})\, .
\end{equation}
The leading term $1/\sqrt{2}$ matches with the expected rate $1/\sqrt{d-2}$ for a 4d emergent string limit. For the subleading term notice that the term $\log[s]$ always dominates, which means in particular that  $|\nabla \Lambda_s|/\Lambda_s$ must approach the bound $1/\sqrt{2}$ from above leading to the characteristic asymptotic behavior that we observed already in figure \ref{fig:T2slice}.

To close our discussion of this example, let us briefly discuss the point where $|\nabla \Lambda_s|/\Lambda_s>1/\sqrt{2}$ is maximized, which lies between the desert point $t=e^{2\pi i /3}$ and the infinite distance point $t=i\infty$. In the regime of large $s \gg 1$ we can use the approximation \eqref{eq:T2approx} to find the position of the maximum as
\begin{equation}\label{eq:T2max}
    s_{\rm max} = \frac{3 \Gamma(\frac{1}{3})^6 e^2}{32\pi^4} e^{N_{\rm des}/6}\, .
\end{equation}
By increasing the $N_{\rm des}>1$ we can thus increase $s_{\rm max}$ and thereby move this maximum further towards the infinite distance point. We can also give a prediction for the amount by which we violate the bound $1/\sqrt{2}$: plugging \eqref{eq:T2max} back into \eqref{eq:T2approx} we find that the maximum value is given by
\begin{equation}\label{eq:T2grad}
    \frac{|\nabla \Lambda_s|}{\Lambda_s}\bigg|_{t=\frac{1}{2}+is_{\rm max}} = \frac{1}{\sqrt{2}}+\frac{32\pi^3}{\sqrt{2}\Gamma(\frac{1}{3})^6e^2}e^{-N_{\rm des}/6}\, .
\end{equation}
Thus as we increase the number of species $N_{\rm des}$ at the desert point we find that the violation amount falls off exponentially. Hence, to infer the value of the constant $c$ appearing in \eqref{QGbound} we first need to know the number of species $N_{\rm des}$ at the desert point.

\subsubsection*{Example 2: Quintic \texorpdfstring{$X_5(1^5)$}{X5}}
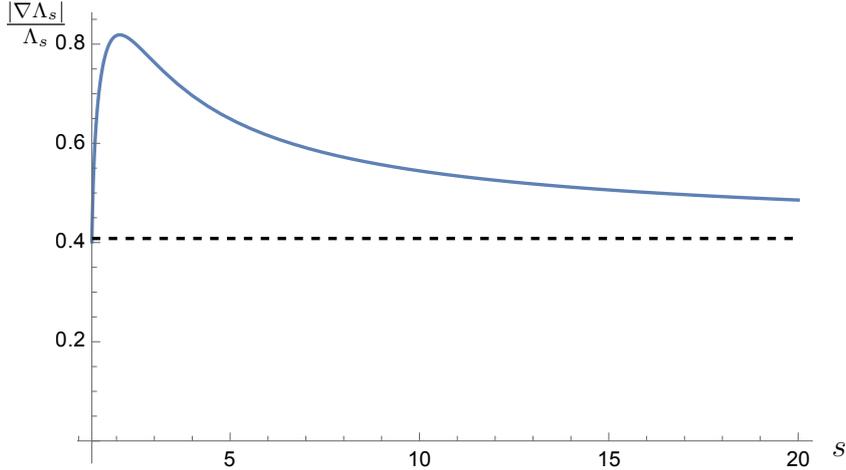
\begin{figure}[!t]
\begin{center}
\includegraphics[width=10cm]{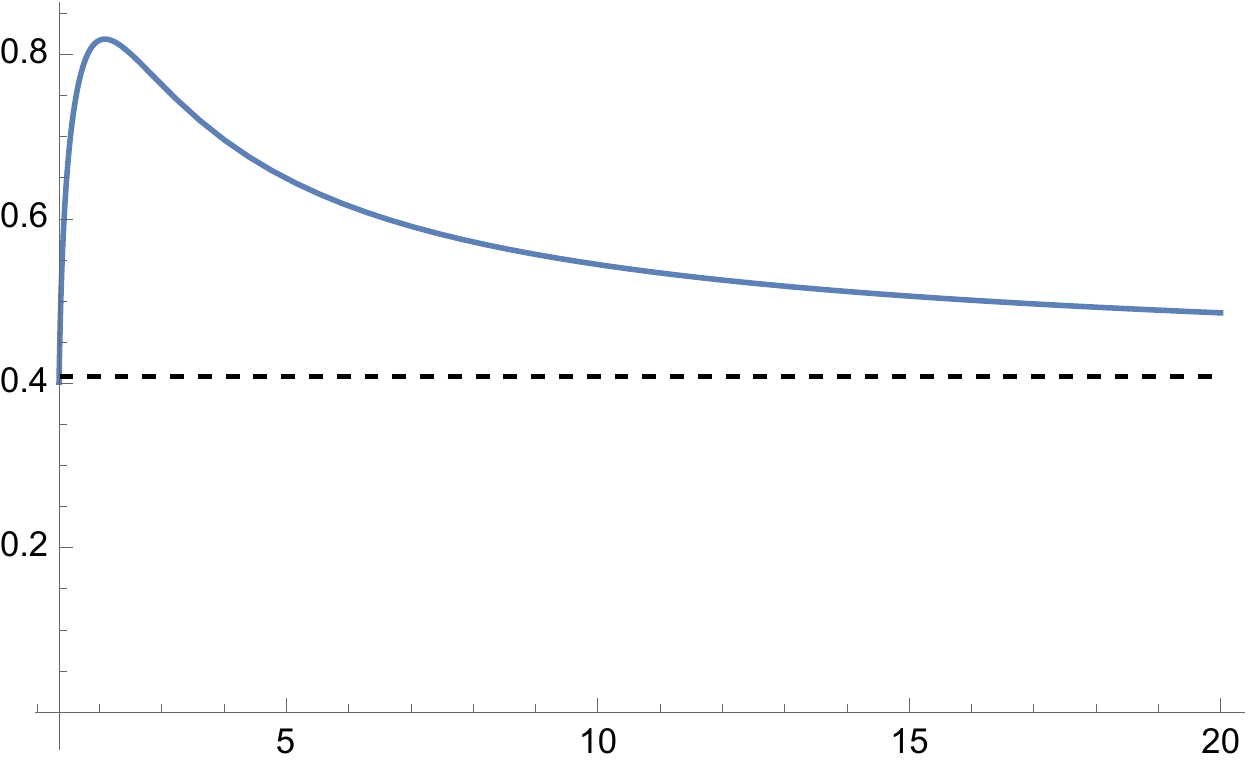}
\end{center}
\begin{picture}(0,0)
\put(80,195){$\frac{|\nabla \Lambda_s|}{\Lambda_s}$}
\put(390,35){$s$}
\end{picture}\vspace*{-0.8cm}
\caption{\label{fig:X5slice} Plot of the gradient $|\nabla \Lambda_s|/\Lambda_s$ for a slice in the moduli space of the quintic. We consider the line between the conifold and the large volume point parametrized by the coordinate $t=a+is=\log[x]/2\pi i$ (such that the large volume point is at $s=\infty$) and setting the axion to $a=0$. The curve corresponds to the ratio $|\nabla \Lambda_s|/\Lambda_s$ for $N_{\rm des} = 1$, while the dashed line indicates the asymptotic bound $|\nabla \Lambda_s|/\Lambda_s = 1/\sqrt{6}$.}
\end{figure}

In this section we consider the quintic threefold as an example for a large volume limit. We use a coordinate $x$ on the moduli space such that the large volume point lies at $x=0$, the conifold point at $x=5^{-5}$ and the Landau-Ginzburg point at $x=\infty$. The genus-one free energy for the quintic is given by
\begin{equation}
    F_1 = \frac{31}{3} K - \frac{1}{2}\log[G_{x\bar{x}}]+\log|x^{-\frac{31}{12}}(1-5^5x)^{-1/12}|\, .
\end{equation}
In \cite{vandeHeisteeg:2022btw} the Landau-Ginzburg point was identified as the desert point where $F_1$ is minimized. Here we want to focus on the behavior of $\Lambda_s$ close to the large volume point following the general discussion in the previous section should correspond to a one-dimensional decompactification limit. To illustrate this behavior we have provided a plot of $|\nabla \Lambda_s|/\Lambda_s$ in figure \ref{fig:X5slice} along a real line towards the large volume point. We find that the gradient $|\nabla \Lambda_s|/\Lambda_s$ reaches a maximum before it descends to the asymptotic value $1/\sqrt{6}$ associated to this decompactification from four to five dimensions.

As discussed in the general case, this behavior of $|\nabla \Lambda_s|/\Lambda_s$ can be explained by the sign of the correction term proportional to $\log[s]$. For the quintic we can expand $F_1$ around this large volume point as
\begin{equation}
    F_1 \to  \frac{50}{12} 2\pi s -30 \log s +\mathcal{O}(s^0)\, ,
\end{equation}
where we suppressed an additive constant and further exponential corrections. Notice that for the quintic $h^{1,1}=1$, $h^{2,1}=101$ such that the coefficient of $\log[s]$ agrees with the general expectation in \eqref{betas}. This again signals that the species scale gradient $|\nabla \Lambda_s|/\Lambda_s$ descends to the asymptotic value associated to the limit --- in this case $1/\sqrt{6}$ --- from above.

\subsubsection*{Example 3: The bicubic \texorpdfstring{$X_{3,3}(1^6)$}{X33}}
\begin{figure}[!t]
\begin{center}
\includegraphics[width=10cm]{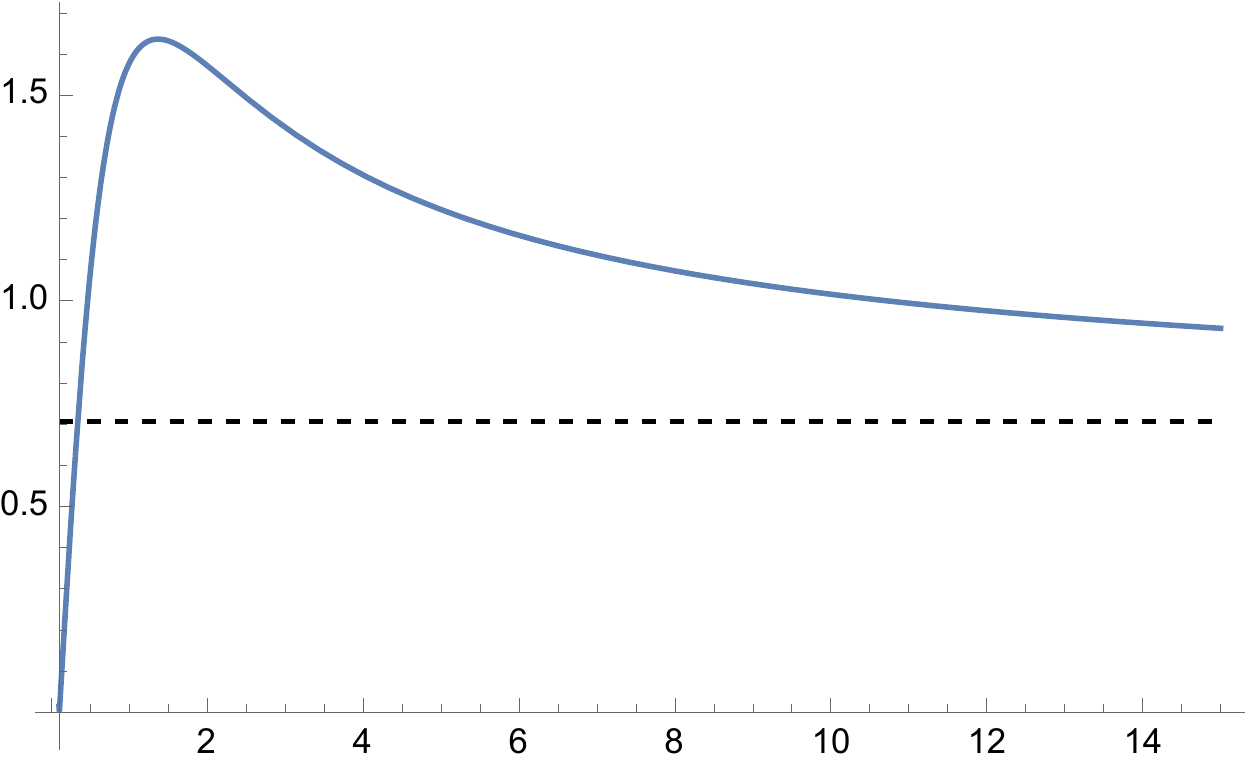}
\end{center}
\begin{picture}(0,0)
\put(80,195){$\frac{|\nabla \Lambda_s|}{\Lambda_s}$}
\put(390,35){$s$}
\end{picture}\vspace*{-0.8cm}
\caption{\label{fig:X33slice} Plot of the gradient $|\nabla \Lambda_s|/\Lambda_s$ in the moduli space of the bicubic. We consider the line between the conifold and the K-point parametrized by $t=a+is=-\log[x]/2\pi i$ (such that the K-point is at $s=\infty$) and setting $a=0$. The curve corresponds to the ratio $|\nabla \Lambda_s|/\Lambda_s$ for $N_{\rm des} = 1$, while the dashed line indicates the asymptotic bound $|\nabla \Lambda_s|/\Lambda_s = 1/\sqrt{2}$ for the emergent string limit. We also note that we start from the desert point around $t=0.103437 i$ where $\Lambda_s$ is maximized and $|\nabla \Lambda_s|/\Lambda_s = 0$, matching with $x_{\rm des}=1.91538$ found in \cite{vandeHeisteeg:2022btw}.}
\end{figure}
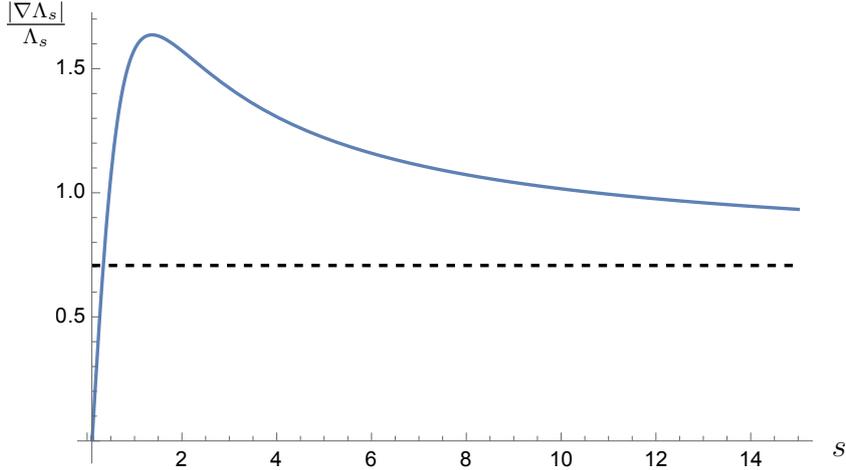

For our final example we consider the bicubic $X_{3,3}$. We parametrize its moduli space such that the large volume point lies at $x=0$, the conifold point at $x=3^{-6}$ and a K-point at $x=\infty$. The purpose of this example is to study the behavior of $|\nabla \Lambda_s|/\Lambda_s$ near the K-point. This singularity corresponds to an emergent string limit. Before we studied such an infinite distance limit for the $(\rm{K3}\times T^2)/\mathbb{Z}_2$  compactification, and now we can investigate the behavior of the species scale in a more non-trivial setting. The genus-one free energy of the bicubic reads
\begin{equation}
    F_1 = 8K - \frac{1}{2} \log[G_{x \bar x}] + \log\big|x^{-\tfrac{11}{2}}(1-3^6 x)^{-1/12}\big|^2\, .
\end{equation}
In \cite{vandeHeisteeg:2022btw} the desert point of this example was identified in between the conifold point and the K-point at $x_{\rm des}=1.91538$. Here we want to focus on the behavior towards the K-point. For illustration we have provided a plot of $|\nabla \Lambda_s|/\Lambda_s$ in figure \ref{fig:X33slice} between this desert point and the K-point in the coordinate $t=a+is=-\log[x]/2\pi i$ (where we set the axion $a=0$). We find, similar to the $(K3\times T^2)/\mathbb{Z}_2$ example that $|\nabla \Lambda_s|/\Lambda_s$ reaches a maximum before it descends to the asymptotic value $1/\sqrt{2}$ of the 4d emergent string limit.

Similar to the other two examples, we can again trace this asymptotic behavior of $|\nabla \Lambda_s|/\Lambda_s$ back to a logarithmic correction $\log[s]$ to $F_1$. By expanding $F_1$ around the K-point for large $s\gg 1$ we find that
\begin{equation}\label{eq:X33approx}
    F_1 \to 4\pi s  - 7 \log[s]+ \mathcal{O}(s^0).
\end{equation}
The leading term $4\pi s$ matches with the expected coefficient $4\pi c_2/24$ for a K3-fibration (which has $c_2=24$). The subleading term $\log[s]$, similar to the other two examples, has a negative sign. This signals again that the gradient $|\nabla \Lambda_s|/\Lambda_s$ descends to the asymptotic value associated to the limit --- in this case $1/\sqrt{2}$ --- from above. The coefficient of this term matches with the result \eqref{betas} for emergent string limits by plugging in $h^{1,1}=1$ and $h^{2,1}=73$ (and $h^{1,1}_{\rm exc}=0$) for the bicubic.

\section{Concluding Remarks}\label{sec:conclusions}
In this paper we have argued for an upper bound on the variation of the species scale with respect to massless moduli:
\begin{equation}\label{boundconcl}    
\left|{\Lambda'_s(\phi)\over \Lambda_s(\phi)}\right |^2<{c\over M_{\rm pl}^{d-2}}\, .
\end{equation}
This argument is intrinsically gravitational and it uses the effective action of gravity motivated by the black hole entropy.  It is interesting that this bound holds at all points in the moduli space, and not just asymptotically. We have also argued that this implies that the mass scale of a light tower of states cannot go to zero faster than exponential as we approach the boundaries of moduli space.   The Distance Conjecture, which demands exponential vanishing of the masses with distance in field space as we approach the boundaries, saturates this exponential behaviour.  The bound we have found is the first clear argument where the {\it exponential} behaviour of the species scale and the mass tower appears as a special limiting case.   It would also be interesting to connect this bound with the dS Swampland conjectures \cite{Obied:2018sgi,Ooguri:2018wrx,Bedroya:2019snp} as it seems to have some formal similarities to them.

In addition, we tested the validity of the expectation $c=1/(d-2)$ for the $\mathcal{O}(1)$ constant in \eqref{boundconcl} motivated from the emergent string conjecture. To that end we studied corrections to the slope of $\Lambda_s$ in the asymptotic regimes, and found that in fact $|\nabla \Lambda_s|/\Lambda_s$ can approach its asymptotic value from above, thereby implying that $c$ in \eqref{boundconcl} has to be bigger than $1/(d-2)$.
Given the power-law relation between the mass $m(\phi)$ of the tower of light states and the species scale $\Lambda_s$, it is natural to believe that the observed behavior of $|\Lambda_s'|/\Lambda_s$ reflects that of $|m'/m|$. In that case also $|m'/m|$ could approach its asymptotic value from above. It would be interesting to study this further.

Finally, the finiteness of the black hole entropy was investigated as a motivation for the Distance Conjecture in \cite{Hamada:2021yxy}.  It would be interesting to try to tighten this argument and in particular to obtain the exponential vanishing of masses with the field space distance.  Our arguments in this paper only show this is an upper bound on how fast the mass can fall off as we approach infinity.

\subsubsection*{Acknowledgments} 
We would like to thank Alek Bedroya, Chuck Doran, Zohar Komargodski, Dieter L\"ust, Miguel Montero, John Stout, Irene Valenzuela, Timo Weigand, David Wu, and Kai Xu for interesting discussions and correspondence. The work of CV and MW is supported by a grant from the Simons Foundation (602883,CV) and by the NSF grant PHY-2013858.

\bibliography{papers_Max}
\bibliographystyle{JHEP}

\end{document}